\documentclass[aps,pre,reprint,groupedaddress,a4paper,onecolumn,notitlepage]{revtex4-1}

\usepackage{lmodern,enumerate,natbib}
\usepackage{amsmath,amssymb,amsthm,amsfonts,mathrsfs}
\usepackage[colorlinks=true,linkcolor=blue,citecolor=blue,urlcolor=blue]{hyperref}
\usepackage{epsfig,epstopdf}

\newcommand{\be}{\begin{equation}}
\newcommand{\ee}{\end{equation}}
\newcommand{\bea}{\begin{eqnarray}}
\newcommand{\eea}{\end{eqnarray}}
\newcommand{\ben}{\begin{enumerate}}
\newcommand{\een}{\end{enumerate}}
\newcommand{\bit}{\begin{itemize}}
\newcommand{\eit}{\end{itemize}}

\newcommand{\bert}{\raise-0.45mm\hbox{\Large$\Box$}}	%D'Alembertian

\begin{document}

\title{On the meaning of  ``quantum supremacy'' experiments.}

\author{Robert Alicki}
\email{Robert.Alicki@ug.edu.pl}
\affiliation{ International Centre for Theory
of Quantum Technologies (ICTQT), University of Gda\'nsk, \linebreak 80-308 Gda\'nsk, Poland}
\date{\today}

\begin{abstract}
The recently reported experimental results claiming ``quantum supremacy'' achieved by  Google quantum device are critically discussed.
The Google team  constructed a quantum chaotic system based on  Josephson junction technology which cannot be reliably
simulated by the present day supercomputers. However, the similar ``supremacy'' can be realized  for properly designed micro-mechanical devices, like periodically forced
Duffing oscillator, using the available technology of quartz clocks. It is also reminded that  classical and quantum chaotic systems behave in a similar way. Therefore, in this case,  one should speak rather about
the ``analog supremacy''  than ``quantum supremacy'' what means that even now mechanical analog computers can outperform supercomputers when the computational task 
can be reduced to sampling of ergodic measures generated by chaotic systems.

\end{abstract}


\maketitle

\section{Introduction}
The Google superconducting device  with 53 qubits executing quantum algorithms with 1500 gates is a remarkable achievement of technology and experimental physics \cite{google}. The single-qubit error of $0.16 \%$, two-qubit error of $0.93\%$ and the measurement error of  $3.8 \%$ in comparison with  the average $\sim3\%$ error for the two-qubit device used to check violation of Bell's inequality in 2009 \cite{bell} is also an essential progress. However, the decrease of an error by less than an order of magnitude during the decade suggests that the superconducting qubit technology is reaching its limits in minimizing errors.  Indeed, as shown theoretically in \cite{Qdot} for the case of excitonic degree of freedom in the quantum dot and then confirmed by computations for different degrees of freedom, the error in any quantum dot  single-qubit gate cannot be lower than  $\sim 0.1\%$ \cite{Jacak}.
Because the fundamental and inevitable sources of errors, i.e. electromagnetic and  acoustic noise,  are the same for quantum dots and superconducting systems one can expect also the similar bounds on the minimal errors.

The arguments in favor of ``quantum supremacy'' are based on the plausible hypothesis that the sequence of gates performed by the device can be treated as an evolution of a chaotic quantum open system.
Therefore, in the next section the comparison between quantum and classical chaos is discussed. Finally, the proposal of realistic implementation of a classical device with chaotic evolution is used to illustrate the  ``supremacy of chaotic (classical) dynamical systems over digital computers''.

\section{Classical vs. quantum chaos}

Remarkably, in the early period of activity in the fields of ``quantum chaos'' and ``quantum computations'' the similar arguments were used to stress the supposed fundamental differences between quantum cases and their classical analogs. One used to say that quantum systems cannot be chaotic because the evolution equation (Schroedinger eq.) is linear in contrast to nonlinear classical equations of motion.  Similarly, linearity of quantum equations and linearity of quantum error maps were supposed to be responsible for unusual stability of quantum evolution and feasibility of fault-tolerant quantum computations.

The closer look at the problem supported by numerical simulations of quantum systems with chaotic classical analogs shows a completely different picture - a close correspondence between dynamical behavior of classical chaotic systems and their quantum counterparts (see \cite{AL2007} , \cite{Zyczkowski}, for details and references).  In the absence of the notion of trajectory in phase-space for quantum system natural mathematical tools are provided by entropic parameters, based on the idea of Kolmogorov-Sinai entropy in the theory of (classical) dynamical systems.  Those entropies  characterize  instability of state evolution for chaotic systems perturbed by repeated measurements with a given accuracy (or similarly by random noise). Such  perturbed states produce  probability distributions (density matrices) with exponentially growing essential supports, both for classical and quantum systems. The only difference is that the corresponding entropy, roughly proportional  to the logarithm of the size of this support, grows linearly to infinity for classical systems and saturates at the fixed value for quantum ones. However, the initial slope for the quantum system coincides with the slope determined by the K-S entropy for its classical counterpart.  The saturation of the quantum entropy at the value  $\ln (dim \mathcal{H})$ (or $2\ln (dim \mathcal{H})$, for different definition of entropy), where $\mathcal{H}$ is the Hilbert space if the system  (see Fig.1a), is quite obvious. The  compact $2d$-dimensional phase-space of the volume $\Omega$ is replaced by the Hilbert space of the dimension $dim \mathcal{H}= \Omega/\hbar^d$. Hence the maximal entropy of the state on $ \mathcal{H}$ is  $\ln (dim \mathcal{H})$ (additional factor 2 for the other definition comes from the additional ancillary system), while  the corresponding entropy for continuous classical system can be arbitrarily large due to arbitrarily fine coarse-graining.

\section{Simulation of classical chaotic system}

One can ask the question whether it is possible to construct a classical device exhibiting chaotic evolution which can be controlled and measured  with a reasonable accuracy for  a certain interval of time, while numerical simulations of its trajectories in the same time interval are  not feasible, even for the most advanced supercomputers.  Consider one of the simplest mechanical models exhibiting chaos, namely, periodically forced Duffing oscillator. Its equation of motion 
\begin{equation}
\ddot{x} + \delta \dot{x} - \alpha x + \beta x^3 = \gamma \cos(\omega t),\quad \alpha, \beta, \delta, \gamma, \omega > 0. 
\label{Duffing}
\end{equation}
which can be written in the autonomous  form
\begin{align}
\dot{x} &= y  , \nonumber\\
\dot{y} &=  \alpha x - \beta x^3 - \delta y + \gamma \cos z, \nonumber\\
\dot{z} &= \omega ,
\label{Duffing1}
\end{align}
exhibits chaotic behavior, for the suitable choice of parameters. The possible physical implementation of the Duffing oscillator is presented on Fig.1b. A  micro-mechanical quartz cantilever  with  an attached steel tip and placed between two micro-magnets is well described by the double-well quartic-quadratic potential of the Duffing oscillator. Using the piezoelectric properties of quartz crystals and the  technology applied for quartz clocks one can also implement periodic driving and measurement of the trajectory $x(t)$. The accuracy of the simulation by such a device can be inferred from the accuracy of good quartz clocks. This  is given by the number of  cycles  $N_c \sim 10^6$ after which the information about the initial phase of the oscillator is lost, mainly due to thermal fluctuations of the device parameters. 

To estimate, how many cycles  of periodic forcing can be simulated on the supercomputer one can use the results of ref. \cite{supercomputer}.
Using 1200 CPUs of the National Supercomputer TH-A1, the authors have done a reliable simulation of chaotic solution of Lorenz equation in a rather long interval $0 \leq t \leq 10000$ LTU (Lorenz time unit). 

Assuming that the Lorentz system, 
\begin{align}
\dot{x} &= \sigma (y - z)  , \nonumber\\
\dot{y} &=  Rx - y - xz, \nonumber\\
\dot{z} &= xy - bz ,
\label{Lorentz}
\end{align}
with the same dimension of phase space as the Duffing oscillator \eqref{Duffing1} can provide a generic benchmark, one can argue that the number of cycles for chaotic solutions of Duffing oscillator  which can be simulated on a supercomputer is of the order of $10^4$, which is smaller by two orders of magnitude than the expected range of effective simulations performed by the analog device proposed above.

\begin{figure} [t]
\begin{center}
\includegraphics[width=0.50 \textwidth]{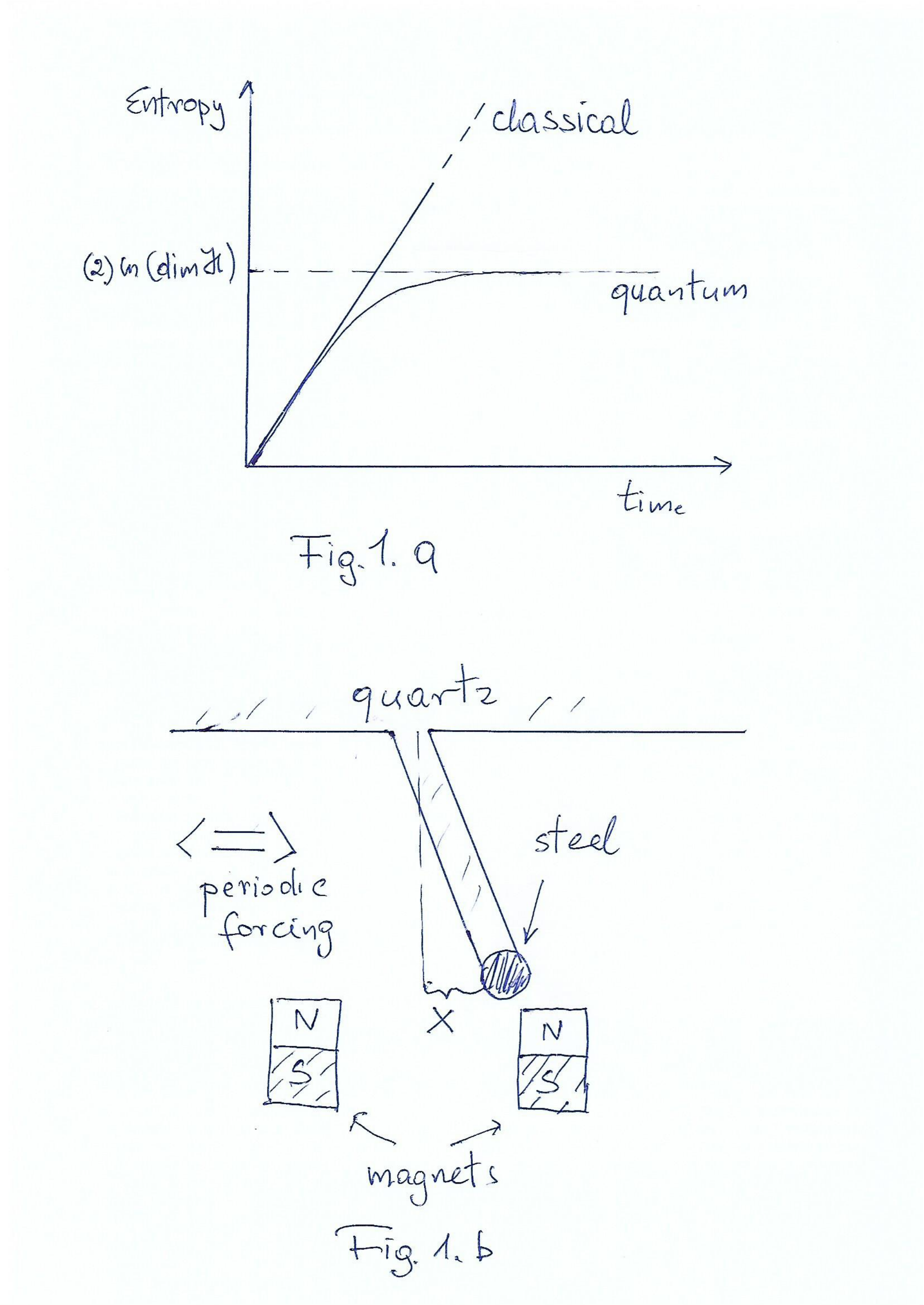}
\end{center}
\caption{\small a) Schematic plots of a dynamical entropy as a function of time  for a chaotic classical system and its quantum counterpart. b) Proposal of a physical implementation of the periodically forced Duffing oscillator.}
\end{figure}

\section{Conclusions}

The proposed analog chaotic device executes the procedure similar to the Google quantum device task, namely sampling of a certain measure generated by  chaotic dynamical system. This ``analog  sampling'' in both cases can be more  efficient than that provided by digital supercomputers. Analog computers with continuous variables were in the past and  seem to be  still  better suited for solving certain specific computational problems. The discussed models supply another arguments for a  not very popular opinion that quantum computers share with classical analog computers all their ups and downs \cite{AL2013}.

\end{document}